# Accuracy Enhancement of Pickett Tunnelling Barrier Memristor Model

Ahmad A. Daoud[1], Ahmed A. Shaaban, Sherif M. Abuelenin

Dept. of Electrical Engineering
Faculty of Engineering, Port Said University
Port Said, Egypt
[1] a.a.daoud@outlook.com

*Abstract*—Titanium dioxide ($TiO_2$) memristors exhibit complex conduction mechanism. Several models of different complexity have been developed in order to mimic the experimental results for physical behaviors observed in memristor devices. Pickett's tunneling barrier model describes the $TiO_2$ memristors, and utilizes complex derivative of tunnel barrier width. It attains a large error in the ON switching region. Variety of research consider it as the reference model for the $TiO_2$ memristors. In this paper, we first analyze the theory of operation of the memristor and discuss Pickett's model. Then, we propose a modification to its derivative functions to provide a lower error and closer agreement with physical behavior. This modification is represented by two additional fitting parameters to damp or accelerate the tunnel width derivative. Also, we incorporate a hard limiter term to limit the tunnel width to its physical extremes 1 nm and 2 nm. We run simulations to test the model modifications and we compare the results to the experimental and original Pickett's model results. The modified model more closely resembles the experimental behavior of $TiO_2$ memristors and potentially enables the memristor to be used as a multilevel memory.

## I. Introduction

In 1971 Leon Chua [1] theorized a fourth fundamental passive circuit element, which he called a memristor (memory-resistor). The memristor is a promising device that could be used in various applications with high packing density due to its nano-scale dimensions. Unlike the theorized memristor, manufactured memristors have complex behavior, and require accurate modeling. The first fabricated memristive device is a Titanium Dioxide ($TiO_2$) based device that was introduced by HP Labs [2] in 2008.

Since the realization of $TiO_2$ memristors, several models have been introduced to capture its behavior; this includes models based on linear [3] and non-linear dopant drift assumption [3-9].

The linear dopant drift mechanism is the simplest and least accurate one. It describes the memristor as two variable resistors connected in series whose values vary with applied excitation field, which causes carriers migration. The drift velocity is constant throughout the entire length of the memristor, also there is no mathematical consideration concerning the limits of physical boundaries. Thus, it does not accurately describe the physical behavior. Hence, a need for a nonlinear dopant drift arises. The nonlinear model adds a window function to the mathematical representation in order to impose a varying drift velocity and resolve the boundary conditions which restrict the change of the state variable to the physical limits of the memristor.

The third mechanism is a tunneling based model as presented by Pickett et al. [10]. It models the memristor as two contacts with a tunnel barrier sandwiched between them. It utilizes Simmon's tunneling equations [11] to control the physical behavior of the memristor. This model shows a very close correlation to the characterizations of practical $TiO_2$ memristors.

In this paper, we investigate the behavior of Pickett tunnel barrier model [10]. Then, we incorporate two scaling factors into the derivative functions of the model and limit the changes of the tunnel barrier width to its physical extremes identified by [10], i.e., 1.1 to 1.9 nm $\pm 0.1$ nm or simply we could consider them as 1 and 2 nm. We verify the advantages of this modification and validate it through comparing the SPICE simulations' results with the experimental results reported in [12]. A good reason for the modification is the potential use of the modified model to represent the memristors in multilevel memories [13] as it resembles the in-between resistances which lay between the maximum and minimum extremes, otherwise the model may confine its utilization to specific applications such as binary memories or switches.

We organize the rest of the paper as follows. The next section studies the theoretical principle of the mathematical formulation of the tunnel barrier model. We present the modification to the derivative of the tunnel barrier width, the modification to original model PSPICE code and the validation of the modified model through simulation tests in the following two sections. The conclusion is provided in the last section.

## II. Tunnel Barrier Model

Pickett et al. [10] provide tunneling and threshold based asymmetric memristive rate of change. Definitions in (1-8) fully describe the model. The memristor tunneling current as in [10] is:

$$i = \frac{j_o A}{\Delta w^2}\left[\phi_I e^{-B\sqrt{\phi_I}} - (\phi_I + e|v_g|)e^{-B\sqrt{\phi_I + e|v_g|}}\right] \quad (1)$$

where $A$ is the barrier cross-sectional area with a mean value of $10^4 nm^2$, and $j_o$, $\Delta w$, $B$ and $v_g$ are represented by

$$j_o = \frac{e}{2\pi h}, \Delta w = w_2 - w_1, B = \frac{4\pi\Delta w\sqrt{2m}}{h},$$
$$v_g = v - R_i i \quad (2)$$

where $e$ is the electron charge, $m$ is the electron mass, $h$ is Planck constant, $v_g$ is the memristor voltage, $v$ and $i$ are the memristor voltage and current, $R_i = 2400\ \Omega$ accounts for the electrode resistance, whereas the expression of $\phi_I$, $w_1$ and $w_2$ are:

$$\phi_I = \phi_0 - e|v_g|\left(\frac{w_1 + w_2}{w}\right)$$
$$-\left(\frac{1.15\lambda w}{\Delta w}\right)ln\left(\frac{w_2(w - w_1)}{w_1(w - w_2)}\right) \quad (3)$$

$$w_1 = \frac{1.2\lambda w}{\phi_0} \quad (4)$$

$$w_2 = w_1 + w(1 - \frac{9.2\lambda}{(3\phi_0 + 4\lambda - 2e|v_g|)} \quad (5)$$

where $\phi_0$ is the barrier height, $w$ is the tunnel barrier width, and $\lambda$ is expressed as:

$$\lambda = \frac{e^2 ln(2)}{4\pi k \varepsilon_0 w} \quad (6)$$

where $k$ is the dielectric constant, and $\varepsilon_0$ is the free space permittivity. The derivative of the tunnel barrier width ($w$) varies by either

$$\frac{dw}{dt} = f_{off}sinh\left(\frac{|i|}{i_{off}}\right) \times$$
$$exp\left(-exp\left(\frac{w - a_{off}}{w_c} - \frac{|i|}{b}\right) - \frac{w}{w_c}\right) \quad (7)$$

for OFF switching (i > 0), or for ON switching (i < 0),

$$\frac{dw}{dt} = -f_{on}sinh\left(\frac{|i|}{i_{on}}\right) \times$$
$$exp\left(-exp\left(\frac{a_{on} - w}{w_c} - \frac{|i|}{b}\right) - \frac{w}{w_c}\right) \quad (8)$$

where $b$, $w_c$, $a_{on}$, $a_{off}$, $i_{on}$, $i_{off}$, $f_{on}$ and $f_{off}$ are fitting parameters.

## III. PROPOSED MODIFIED MODEL

The results obtained by the model in [12] as compared to the experimental results show about 22.5% error in the memristive current in the ON switching and about 9.91% in the OFF switching. This original model makes $w$ to change from 1.2 to 1.8 nm and then back to 1.1 nm and it deviates from the experimental results. The width seems to accelerate to non-desirable value (1.1 nm). Our purpose is to reduce that error by incorporate a damping parameter $k_{off}$ into the derivative of the tunnel barrier width ($dw/dt$) for OFF switching (i > 0) as expressed in (9).

$$\frac{dw}{dt} = f_{off}\ sinh\left(\frac{|i|}{i_{off}}\right) \times$$
$$exp(-exp\left(k_{off}\left(\frac{w - a_{off}}{w_c} - \frac{|i|}{b}\right)\right) - \frac{w}{w_c}) \quad (9)$$

where the parameter $k_{off}$ has the following role; when assigned a positive value lower than unity, it makes the tunnel width derivative more immune against the term (($w - a_{off})/w_c - |i|/b$). It decelerates the exponential increment (lower $dw/dt$ for lower $w$ and higher $dw/dt$ for higher $w$) for OFF switching. Also we assign different values to the parameters $w_c$ and $b$ to mimic the rest of experimental curve and they are provided below in a following section. These new values along with $k_{off}$ help to reshape the tunnel width variation. To maintain the symmetry between the ON and OFF switching functions, we may add a similar parameter $k_{on}$ to (8). For our fitting results, $k_{on}$ will have a value of unity.

To examine the model accuracy, we reproduce the I-V characteristics of the memristor that uses PSPICE and by simulating the circuit used by [12] as shown in Fig. 1a with excitation input shown in Fig. 1b. Fig. 2a shows a comparison between simulated I-V characteristic results and I-V characteristic of the original model reported in [12].

As shown in Fig. 2a, the proposed model more accurately resembles the experimental curve than the original model especially for the ON switching portion of the curve (negative voltage). Fig. 2b shows the tunnel width $w$ ranges from 1.2 nm to 1.8 nm for the modified model while it ranges between 1.1 nm and 1.8 nm for the original model. This could be considered as advantage of suggested modification over the original model especially for the ON switching region characterized by negative voltage as shown in Fig. 2a.

Fig. 2b shows a damping in tunnel width change in the time interval from 0 to 0.7 s. This change is lower than that of the original model as $w$ approaches 1.2 nm. Then $w$ accelerates in the time interval from 0.7 to 1.5 s to compensate for the former damping. This modified change is higher than that of the original one as $w$ approaches 1.8 nm. This behavior is observed as more curvature in the OFF region (positive voltage) as shown in Fig. 2a where the memristor's current and voltage approach about 0.64 mA and 0.82 V respectively. For negative voltage region, Fig. 2b shows that $w$ decelerates in the time interval from 3.5 to 4 s. This change is lower than the original one as $w$ leaving 1.8 nm. A damping in tunnel width change occurs in the time interval from 4 to 6 s. This change is lower than that of the original model as $w$ approaches 1.2 nm. This behavior is observed as a more curvature in the ON region (negative voltage) in Fig. 2a where the memristor's current and voltage approach –0.28 mA and –0.53 V. The two curves of $w$ share the same width change in the interval from 1.5 to 4 s.

We utilize relative root mean square error in (10) to calculate the error between the experimental data and the proposed model based data for the ON and OFF switching.

$$e_{v,i} = \sqrt{\frac{1}{N}\left(\frac{\sum_{n=1}^{N}(v_{m,n} - v_{r,n})^2}{\bar{v}_r^2} + \frac{\sum_{n=1}^{N}(i_{m,n} - i_{r,n})^2}{\bar{i}_r^2}\right)} \quad (10)$$

where $\bar{v}_r$ and $\bar{i}_r$ are the mean values of the reference (experimental) voltage and current respectively, $v_m$ and $i_m$ are the voltage and current observed by the modified model.

In case of ON switching, the hysteresis loop of the presented model more accurately mimics the experimental data producing an error of 8.13 % whereas the tunneling based model presented by [12] produces an error of 22.5 %. Also the hysteresis loop better resembles the experimental result for the OFF switching region. It produces an error of 8.52 % whereas the tunneling model produces an error of 9.91 %.

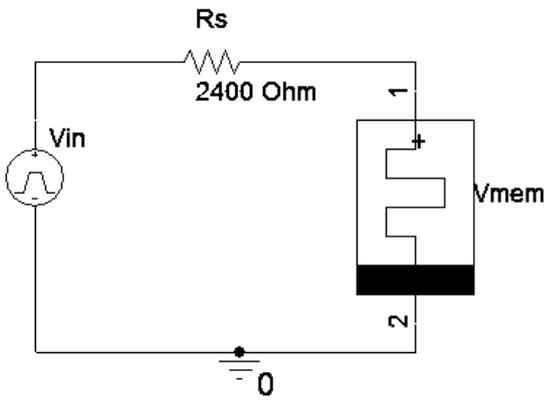

(a)

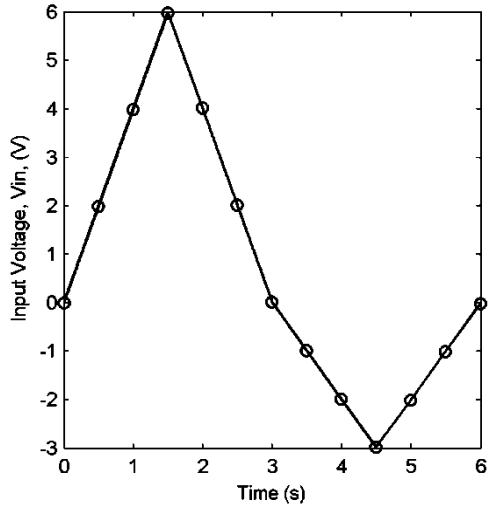

(b)

Fig. 1. Circuit Schematic and input voltage used by [12] for TiO₂ memristor model simulation, (a) Circuit schematic, (b) Input voltage.

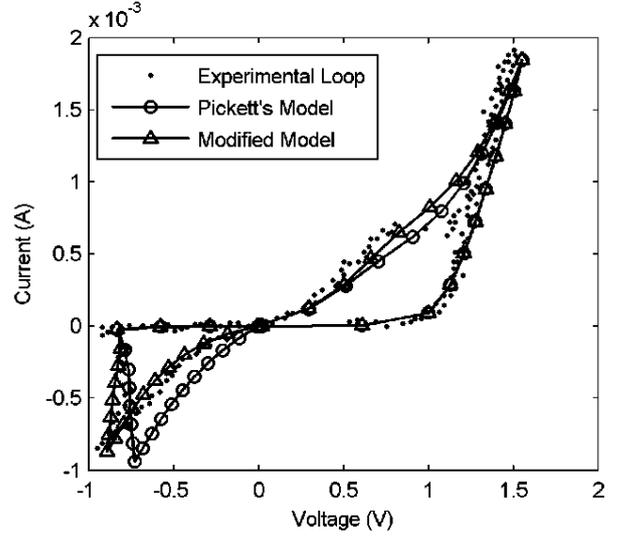

(a)

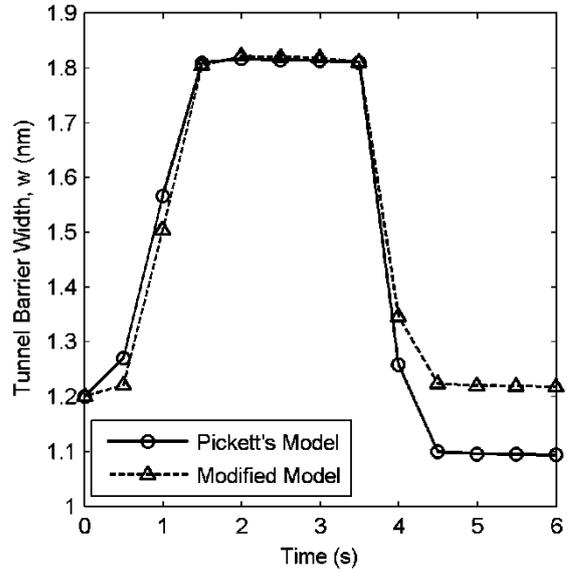

(b)

Fig. 2. Simulation results of modified model that uses the circuit and input voltage shown in Fig. 1 with fitting parameters: $I_{on}$ = 8.9 (μA), $I_{off}$ = 115 (μA), $f_{on}$ = 2 (mm/s), $f_{off}$ = 3.5 (μm/s), $a_{on}$ = 1.8 (nm), $a_{off}$ = 1.2 (nm), $k_{off}$ = 0.5, $w_c$ = 95 (pm) and $b$ = 600 (μA), w is in (nm), (a) Hysteresis I-V loops of experimental results, original tunnel model and that with the proposed modification, (b) Comparison between tunnel barrier widths (w) w. r. to time as provided by the original tunnel model and the proposed modification.

## IV. REPRESENTATION OF PHYSICAL BOUNDARIES IN MEMRISTOR MODEL

Physical boundaries should be limited to no lower than 1 nm and no higher than 2 nm as typical values of the tunnel barrier width [3, 10]. We consider the mathematical limits that represents these boundaries in the proposed model construction i.e., we modify the PSPICE code to include a hard limiter that allows the tunnel width to neither exceed its maximum nor fall down its minimum.

We edit the PSPICE code provided in [12] to present the modified tunnel barrier width derivatives of the tunnel model and add the hard limiter as follows:

```
************************************
.SUBCKT modified_pickett plus minus
.PARAM:
+phio=0.95 Lm=0.0998 w1=0.1261 foff=3.5e–6
+ioff=115e–6 aoff=1.2  ion=8.9e–6 aon=1.8 Th_on=0.0074
+koff1=1 koff2=0.5 kon1=1 kon2=1 b=600e–6 wc=95e–3
+fon=2000e–6
G1 plus internal
value={sgn(V(xp))*(1/V(dw))^2*0.0617*(V(phiI)*exp(–V(B)*V(sr))–(V(phiI)+abs(V(xp)))*exp(–V(B)*V(sr2)))}
Rs internal minus 215
Esr sr 0 value={sqrt(V(phiI))}
Esr2 sr2 0 value={sqrt(V(phiI)+abs(V(xp)))}
Eg xp 0 value={V(plus)–V(internal)}
Elamda Lmda 0 value={Lm/V(w)}
Ew2 w2 0 value={w1+V(w)–(0.9183/(2.85+4*V(Lmda)–2*abs(V(xp))))}
EDw dw 0 value={V(w2)–w1}
EB B 0 value={10.246*V(dw)}
ER R 0 value={(V(w2)/w1)*(V(w)–w1)/(V(w)–V(w2))}
EphiI phiI 0 value={phio–abs(V(xp))*((w1+V(w2))/(2*V(w)))–1.15*V(Lmda)*V(w)*log(V(R))/V(dw)}
********** Edited Terms 1 ************
C1 w1 0 1e–9 IC=1.2
R w1 0 1e8MEG
********** Added Terms 1 ************
Ew w 0 value= {MIN (MAX (V(w1), 1), 2)}
Rw w 0 1T
************************************
Ec c 0 value={abs(V(internal)–V(minus))/215}
Emon1 mon1 0 value={((V(w)–aoff)/wc)–(V(c)/b)}
Emon2 mon2 0 value={(aon–V(w))/wc–(V(c)/b)}
************ Edited Terms 2 **********
Goff 0 w1 value={foff*sinh(stp(V(xp))*V(c)/ioff)*exp(–koff1*exp(koff2*V(mon1))–V(w)/wc)}
Gon w1 0 value={fon*sinh(stp(–V(xp))*V(c)/ion)*(exp(–kon1*exp(kon2*V(mon2))–V(w)/wc))}
************************************
.ENDS
************************************
```

We manipulate three terms in the original PSPICE code [12] to agree with the requirements of neither exceeding nor falling down the typical values of tunnel barrier width *w*. We alter four code lines: replace the node w with w1, and add a new two lines containing a new definition of w.

We run test simulations which Fig. 3 and Fig. 4 shows their results to prove that the memristor which uses the modified model has a width which could neither exceeds 2 nm nor falls down 1 nm. We provide a circuit schematic for the modified model PSPICE code in Fig. 5 that shows the original circuit elements and those added to overcome the deficiencies of the model in [12]. Fig. 3 shows that the tunnel width significantly changes during the rise time of the applied voltage and it seems to remain fixed for DC voltage. Whereas, Fig. 4 shows that the tunnel width changes with the voltage wherever the voltage rises, falls or remains constant. This could be due to the asymmetric ON and OFF switching operations characterized by two asymmetric window functions.

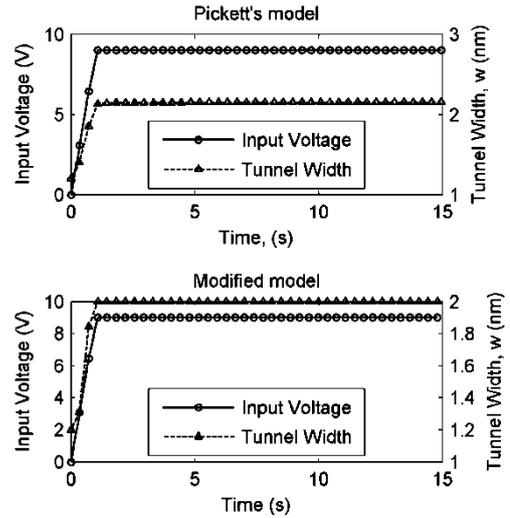

Fig. 3. A simulation test checks whether Pickett's model and the modified one exceed the typical value of highest tunnel barrier (*w* = 2 (nm)) or not. The applied voltage rises form 0 to 9 V for the time interval from 0 to 1 s.

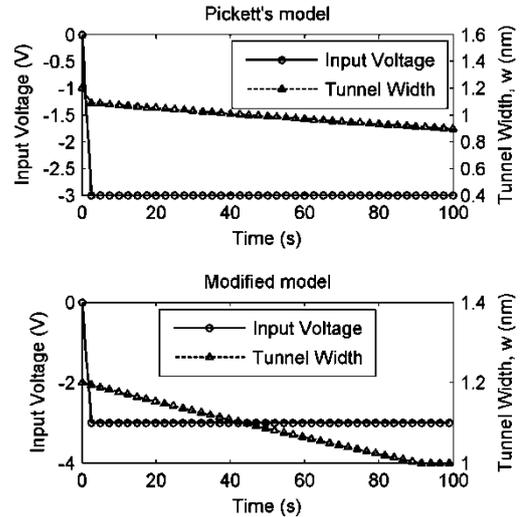

Fig. 4. A simulation test checks whether Pickett's model and the modified one fall down the typical values of lowest tunnel barrier (*w* = 1 (nm)) or not. The applied voltage rises form 0 to –3 V for the time interval from 0 to 1 s.

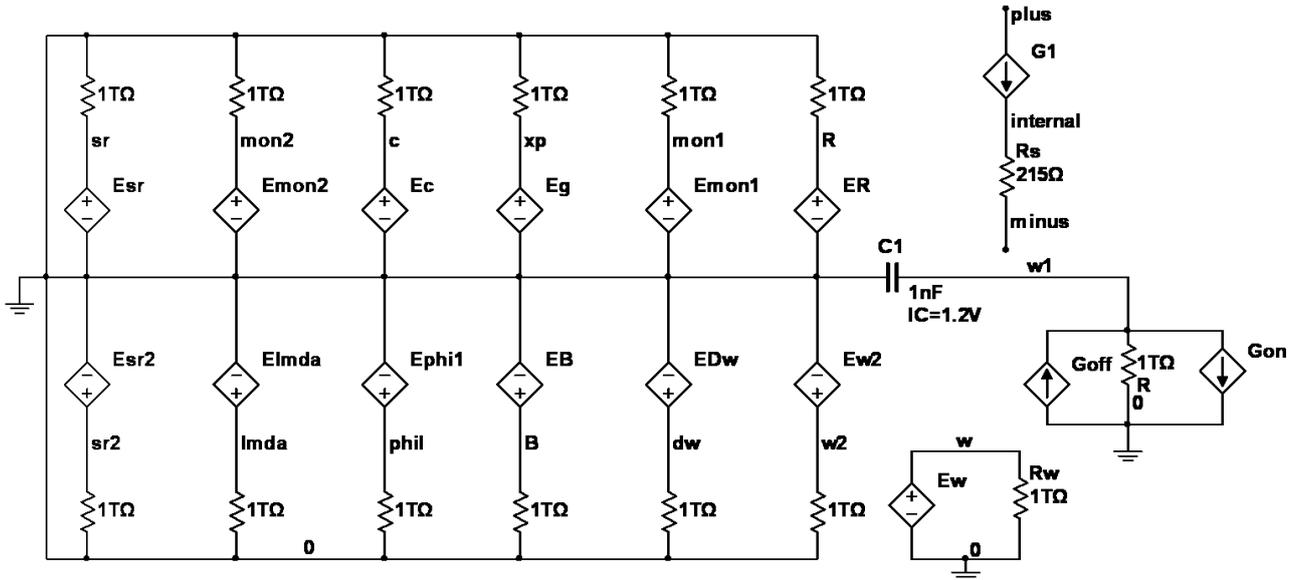

Fig. 5. PSICE sub-circuit as originally coded in [12] and edited in this paper.

## V. CONCLUSION

The modified model provides much better agreement with physical $TiO_2$ memristor behavior as we dampen the fast change of the tunnel barrier width $w$ in the original Pickett's model and reassign two fitting parameters to new values. The $TiO_2$ memristor exhibits 22.5% and 9.91% overestimations in the memristive response when driven with negative and positive voltages respectively as modeled by Pickett tunnel barrier model while the modification to Pickett's model derivative functions shows only 8.13% and 8.52% overestimations. Hence, this could be useful for multilevel memories. Also the modified model could adequately resolve the boundary issues if the memristor is driven to its extremes as it could neither exceed the maximum tunnel width nor fall down the minimum tunnel width unlike the original model that could do under the test voltage. The fit test validates the results and shows a good modified model.